\documentclass[preprints,article,accept,oneauthor,pdftex]{Definitions/mdpi}

\newcommand{\eqb}{\begin{equation}}
\newcommand{\eqe}{\end{equation}}

\firstpage{1} 
\pubvolume{xx}
\issuenum{1}
\articlenumber{5}
\pubyear{2020}
\copyrightyear{2020}

\history{Received: date; Accepted: date; Published: date}

\Title{Emission Mechanisms of Fast Radio Bursts}

\Author{Yuri Lyubarsky 
}

\AuthorNames{Firstname Lastname, Firstname Lastname and Firstname Lastname}

\address{Physics Department, Ben-Gurion University, P.O.B. 653, Beer-Sheva
84105, Israel; e-mail: lyub@bgu.ac.il}

\abstract{Fast radio bursts (FRBs) are recently discovered mysterious single pulses of radio emission, mostly coming from cosmological distances ($\sim 1$ Gpc). Their short duration, $\sim 1$ ms, and large luminosity evidence coherent emission. I review the basic physics of coherent emission mechanisms 
proposed for FRBs. In particular, I discuss the curvature emission of bunches, the synchrotron maser, and the emission of radio waves by variable currents in the course of magnetic reconnection. Special attention is paid to magnetar flares as the most promising sources of FRBs. Non-linear effects are outlined that could place bounds on the power of the outgoing radiation. 
}


\begin{document}

\section{Introduction}

The discovery of fast radio bursts (FRBs) has inspired renewed interest of the astrophysical community in coherent emission mechanisms. An extremely high brightness temperature of the observed emission implies that the emission mechanism is coherent, i.e., a large number of particles must emit the radio waves in phase. Half a century ago, this sort of consideration led to conclusion that the pulsar radio emission is coherent. 
However, a theory of pulsar radio emission has not yet been developed. Therefore, it is not surprising that a pessimistic view of the theoretical study of coherent emission processes is common. 

Under astrophysical conditions, coherent emissions are typically generated in fully ionized plasma (a marked exception is the molecular line masers). The plasma is a medium with a long-range interaction, which, in principle, enables the coherent motions of large ensembles of particles. However, the same long-range interaction strongly affects both the emission and the propagation of electromagnetic waves. Therefore, coherent emission is a collective plasma process that should be described in the language of plasma physics.
Unfortunately, this branch of physics is not popular among
the astrophysical community. Two views are prevalent. According to the first view, 
this is an obscured and untrustworthy field, so there is no chance of any progress. The second view is the opposite (but, in some sense, closely related): coherent emission could be simply described by 
formulas from the Jackson's textbook \citep{Jackson_book}, into which the charge $Ne$ could be substituted with a large enough $N$.

This review advocates for a more insightful approach. The theory of collisionless plasmas supplemented by numerical simulations has already provided spectacular progress in our understanding of astrophysical shock waves, magnetic reconnection, and particle acceleration \citep{Treumann_Baumjohann15,Sironi_etal15,Kagan_etal15,Pelletier_etal17}. The coherent emission from the Sun and planetary magnetospheres has also become a mature field \citep{Melrose17}. The situation with mechanisms of pulsar radio emission looks disappointing but the reason for this is rather specific: it is not that the processes in the magnetospheric plasma defy analysis---the mechanisms of the plasma production and the very structure of the plasma flows in pulsar magnetopsheres remain uncertain, which prevents the confrontation of the theory and observations \citep{Lyubarsky08a}.
 Therefore, there is no reason to halt efforts to resolve the enigma of FRBs. 

This review is an attempt to present, at a basic level, the physics of collective plasma emission mechanisms applied to FRBs. When dealing with radiation processes in plasma, a sensible line of action is to address three basic questions:
\begin{enumerate}[leftmargin=*,labelsep=4.9mm]
\item	What electromagnetic waves are supported by plasma at the assumed conditions?
\item	How could these waves be excited?
\item	Could they escape the system as radio waves?
\end{enumerate}
Any reasonable model has to provide self-consistent answers to these questions. I try to show how this could be achieved in particular cases relevant to FRBs. 

Only basic observational facts are used in this review. First, the theoretical model should explain the $\sim 1$ ms burst duration and a wide range of emitted isotropic energies, from $\sim10^{35}$ to $\sim10^{43}$ erg. Another basic fact is that FRBs typically exhibit strong linear polarization, sometimes reaching 100\%. For a comprehensive analysis of observations, the reader is referred to recent reviews \citep{Cordes_Chatterjee19,Petroff_etal19}. A concise summary of  observational data, as well as the description of theoretical models is given in ref.\ \citep{Zhang20}. 

The remainder of this paper is structured as follows: In Section 2, I describe the coherent radiation mechanisms discussed in the context of FRBs. Magnetars are considered the main candidate source of FRBs. In Section 3, I outline the physics of magnetar magnetospheres and magnetar flares, and discuss how the powerful coherent radiation may be produced by these flares. In Section 4, I describe non-linear processes that could affect the propagation of high intensity radio waves and even prevent the escape of radio emission from the source.

\section{Coherent Radiation Mechanisms}

\subsection{Brightness temperature}
To estimate how many coherently emitting particles are minimally required to produce the observed emission, let us estimate the typical brightness temperature of FRBs. The brightness temperature of a radiation source, $T_b$, is defined such that the radiation intensity of the source at the frequency $\nu$ is presented according to the Rayleigh-Jeans law:
 \begin{equation}
 I_{\nu}=\frac{2k_BT_b\nu^2}{c^2},
 \end{equation}
where $k_B$ is the Boltzmann constant, and $c$ the speed of light. The brightness temperature typically depends on the frequency.

If the source is at rest, the observed frequency and intensity are the same as those in the source. For an isotropic source, the observed spectral flux is:
\begin{equation}
 F_{\nu}=I_{\nu}\Omega\sim\frac{a^2}{D^2}I_{\nu},
\end{equation}
where $\Omega$ is the solid angle subtended by the source, $a$ is the size of the source, and $D$ is the distance. The characteristic spectral flux of FRBs is of the order of 1 Jy; the duration of the bursts, $\tau$, is of the order of 1 ms. The size of the source is limited by $a\le \tau c$, which immediately implies a low estimate of the brightness temperature:
 \begin{equation}
 T_b\sim\frac{F_{\nu}D^2}{k_B\tau^2\nu^2}=7\cdot
 10^{35}\frac{F_{\nu,\rm Jy}D^2_{\rm Gpc}}{\tau^2_{\rm ms}\nu^2_{\rm GHz}} \,\rm K.
 \end{equation}
The indexes like Jy and Gpc indicate the units in which the corresponding quantities are measured. Below, I also employ the shorthand notation $q_x = q/10^x$ in cgs units, e.g., $L_{44}=L/(10^{44}\rm erg/s)$.

In a moving source, the brightness temperature is related to the intensity in the source frame, $I'_{\nu'}=2k_BT_b\nu'^2/c^2$, where the prime means that the corresponding quantities are measured in the source frame.
If the source moves toward the observer with the Lorentz factor $\Gamma\gg 1$, the lower limit on the observed variability time significantly decreases due to the transit time effect:
\begin{equation}
 \tau_{\rm obs}
 \ge\frac{a'}{c\Gamma}.
\label{tau_duration}\end{equation}
To find a relation between the measured spectral flux and the brightness temperature of the source, take into account
that the total luminosity of the source,
\begin{equation}
 L'\sim a'^2I'_{\nu'}\Delta\nu'\sim
 \frac{a'^2k_BT_b\nu'^2\Delta\nu'}{c^2},
\end{equation}
is the relativistic invariant, $L=L'$.
Then, the observed spectral flux is estimated, accounting for the relativistic beaming, as:
\begin{equation}
 F_{\nu}\sim\frac{L}{\Gamma^2D^2\Delta\nu}\sim\frac{a'^2k_BT_b\nu'^2}{c^2D^2\Gamma^2}\frac{\Delta\nu'}{\Delta\nu}.
\end{equation}
Now, the Lorentz transform of the frequency, $\nu\sim\Gamma\nu'$, together with the upper limit on the source size (\ref{tau_duration}), yields the low limit on the brightness temperature:
\begin{equation}
 T_b\sim\frac{F_{\nu}D^2}{k_B\tau^2\nu^2\Gamma}=
 7\cdot 10^{35}\frac{F_{\nu,\rm Jy}D^2_{\rm Gpc}}{\tau^2_{\rm obs, ms}\nu^2_{\rm GHz}\Gamma} \,\rm K.
\end{equation}

The brightness temperature for an ensemble of incoherently emitting particles cannot exceed the temperature of the ensemble, understood as the characteristic energy of particles expressed in degrees. This immediately implies that the emission mechanism of FRBs should be coherent, i.e., a large number of particles must emit radio waves in phase. 
Assuming that the emitting particles are electrons, 
 we estimate the minimal number of coherently radiating particles as:
\begin{equation}
 {\cal N}\sim\frac{k_BT_b}{m_ec^2}\sim
 10^{26}\frac{F_{\nu,\rm Jy}D^2_{\rm Gpc}}{\tau^2_{\rm obs, ms}\nu^2_{\rm GHz}\Gamma}.
 \label{Nminimal}\end{equation}
If the electrons in the source are relativistic, the required number of particles is smaller by the characteristic Lorentz factor of electrons, but in any reasonable case, $N$ remains tremendously large. This implies that the FRB emission is definitely coherent. The question is what causes the concerted motion of a macroscopic number of particles. 

The simplest assumption is that bunches of charged particles are somehow formed in the system, with each bunch emitting as a single entity. Then, two questions immediately arise: what is the bunching mechanism and what are the radiation properties of these bunches? 

\subsection{Curvature Emission of Bunches}

At the very early stage of pulsar studies, particles were found to emit in the radio band if they move with the Lorentz factors $\Gamma\sim 100$ along curved magnetic field lines in the neutron star's magnetosphere. Since then,
the popular view is that bunches of charged particles are somehow formed in the magnetosphere, and that pulsar radio emission could be attributed to the curvature emission of these bunches. Recently, the same model was applied to FRBs 
\citep{Falcke_Rezzolla14,Cordes_Wasserman16,Dai_etal16,Kumar17,Gisellini_Locatelli18,Katz18,Yang_Zhang18,Lu_Kumar18,Wang_etal19,Kumar_Bosnjak20,LuKumarZhang20,Wang_etal20,Yang_etal20}. 

The formation of bunches is typically attributed to the electrostatic plasma waves excited by the two-stream instability. All the available estimates are based on the theory developed at the assumption that the energy of perturbations is small as compared with the energy of the plasma. However, the electrostatic energy of the assumed bunches enormously exceeds the plasma energy. Namely, for the emission to be coherent, the size of the bunch in the comoving frame could not be larger than $c/\omega'$, where $\omega'=2\pi\nu/\Gamma$ is the frequency in the frame of the bunch. Using the estimate (\ref{Nminimal}) for the minimal number of electrons in the bunch, we find the electric potential in the bunch $V'\sim {\cal N}e\omega'/c$. Then the ratio of the electrostatic to the plasma energy is found as
 \begin{equation}
     \epsilon\sim \frac{eV'}{mc^2}\sim 10^{9}\frac{F_{\nu,\rm Jy}D^2_{\rm Gpc}}{\tau^2_{\rm obs, ms}\nu_{\rm GHz}\Gamma_2^2}.
 \end{equation}
In this case, the results of the linear theory are completely irrelevant. The energy of plasma waves could not exceed the plasma energy because due to non-linear effects, it efficiently goes to heat. One can speculate that the plasma is so hot that $\epsilon$ does not exceed unity. However, the properties of both the plasma  waves and the two-stream instability in a relativistically hot, strongly magnetized plasma differ drastically from those in the cold plasma \citep{Lominadze_Mikhailovski79,Arons_Barnard86}. No attempts were made to analyze the formation of the bunches at realistic conditions.

Note that $\epsilon$ represents the repulsive electrostatic potential in units of the electron rest energy. This enormous number clearly shows how difficult is to form the bunches and to keep them against destruction. Unless physical mechanisms capable of formation and maintenance of the bunches will be elaborated,  models assuming the existence of such bunches 
should be considered as highly speculative.

In the papers quoted above, 
the emission of the bunches was estimated by applying standard formulas of electrodynamics in a vacuum. The characteristic frequency is estimated as:
\begin{equation}
 \omega_c\sim\frac{c\Gamma^3}{R_c},
 \label{curv_freq}\end{equation}
whereas the emission power of a bunch is assumed to be:
 \begin{equation}
 P=\frac{2q^2c\Gamma^4}{3R_c^2},
 \label{curv_power}\end{equation}
 where $R_c\sim 10^7-10^9$ cm is the curvature radius of the magnetic field line and $q={\cal N}e$ is the charge of the bunch. The demand that the emitted frequency is in the radio band implies $\Gamma\sim 100$. Choosing an appropriate $\cal N$, we find the required power. 
 
Even leaving aside the question of how the bunches are formed, we cannot
ignore the fact that their radiation properties are strongly affected by the plasma.
The density of this plasma could not be much less than the particle density in the bunches therefore  the minimal comoving plasma density is:
\begin{equation}
 N'\sim\frac{\omega'^3\cal N}{c^3}\sim
 10^{16}\frac{F_{\nu,\rm Jy}D^2_{\rm Gpc}\nu_{\rm GHz}}{\tau^2_{\rm obs, ms}\Gamma^4_2}\,\rm cm^{-3}.
\end{equation}
In this case, the plasma
frequency, $\omega'_p=(4\pi e^2N'/ m_e)^{1/2}$, is well above the frequency of the emitted waves:
\begin{equation}
 \frac{\omega_p'}{\omega'} \sim 10^5 
 \frac{F_{\nu,\rm Jy}^{1/2}D_{\rm Gpc}}{\tau_{\rm obs, ms}\nu_{\rm GHz}^{1/2}\Gamma_2}.
\end{equation}
Waves with frequencies below the plasma frequency could propagate in the highly magnetized plasma of a neutron star's magnetosphere , see, e.g., \citep{Arons_Barnard86}. However, their emission cannot be described by formulas of vacuum electrodynamics. 

The curvature emission of a point charge at the condition $\omega'_c\ll\omega'_p$ was calculated in ref.\ \citep{Gil_etal04}. It was  found that the characteristic radiation frequency is the same as that in the vacuum case, Eq. (\ref{curv_freq}), but the emission power is strongly suppressed:
\begin{equation}
 P=K_0\frac{q^2c\Gamma^4}{R_c^2}\left[\frac{\omega'_c}{\omega'_p}\left(1-\frac{\Gamma^2_p}{\Gamma^2}\right)\right]^2,
\end{equation}
where $\Gamma$ is the Lorentz factor of the charge, $\Gamma_p$ is the Lorentz factor of the plasma, $K_0\approx 0.1$, and the prime means that the corresponding quantities are measured in the frame comoving with the plasma. If the charge moves
together with the plasma, $\Gamma=\Gamma_p$, it does not radiate at all because it is completely shielded in this case. Note that in this calculation, the plasma was assumed to be cold. Hot plasmas completely shield charges moving with a velocity smaller than the electron thermal velocity \citep{Krall1973}. Therefore, in a relativistically hot plasma, the curvature emission could be expected to be completely suppressed for a wide range of $\Gamma$. If the charge moves with respect to the plasma such that its velocity is out of the range of the plasma thermal velocities, it could emit but the emission rate would be suppressed by a factor $(\omega'/\omega'_p)^2\sim 10^{-10}$ compared with the vacuum emission rate (\ref{curv_power}). 

Of course, this highly idealized result could be modified by considering additional effects, such as non-point-like charge distribution, hot plasma, etc. However, this is a burden on the authors of the model to present a self-consistent description of all relevant physical processes. The above example just shows that the plasma effects could not be
ingnored in the emission process. It is worth noting that in pulsars, the plasma frequency is comparable with the curvature frequency, and the electrostatic energy of the assumed bunches is not too large in typical cases. Therefore, in this case, the model of the curvature emission could not be certainly ruled out, even though speculations of the origin of bunches are not convincing. In all the available versions of the coherent curvature emission model for FRBs, both the formation of bunches and their emission is described using theoretical results obtained at the conditions by far violated in the model.  Therefore for a while, this model could not be considered physically consistent, so that the discussion of any details of the model is premature. 

\subsection{Masers, Some Preliminaries}

The standard explanation of maser action is based on the notion of induced emission introduced by Einstein in his analysis of the quantum radiation transitions. A photon with energy matching the energy difference between two energy levels could not only be absorbed by the system, causing a transition of an electron from the lower to the upper level, but could also cause an electron in the upper level to jump downward, emitting one more photon with the same energy and direction. Therefore, if the population of the upper level exceeds that of the lower level, a seed radiation beam is exponentially amplified. 

The above quantum picture is well known and is, in principle, correct, even though it could be sometimes misleading, as with any purely qualitative picture. But it is worth stressing that we deal with classical systems, so even though the classical view is paradoxically less vivid than the quantum view, quantitative theory may be developed using only classical plasma physics. 

According to the classical picture, maser action occurs in resonantly unstable systems. A seed wave satisfying the resonance conditions modulates the unstable plasma, triggering currents that emit in phase with the seed wave, thus amplifying it. In this case, the maser emission could also be attributed to charged bunches formed in the course of the modulation process. However in this approach, both the formation of the bunches and their emission are considered self-consistently, accounting for the dispersion properties of the medium.
 
\subsection{Synchrotron Maser} 
Charged particles rotate in a magnetic field; therefore, an inverse population in the magnetized plasma could be simply imagined as a ring in the particle momentum space. Before considering how such a distribution could be formed in nature, a couple of subtle points must be discussed.

First, maser action is impossible in an infinite system of equidistant energy levels, like the system of non-relativistic Landau levels \footnote 
{The cyclotron maser works only due to relativistic corrections to Landau levels \citep{Melrose17}}. Even if the $n$th level is overpopulated, the resonant photon could trigger both the transition $n\to n-1$ (induced emission) and $n\to n+1$ (absorption). The absorption rate is larger, so the resonant photons are always absorbed in this case. In particular, maser action is impossible in a system of linear oscillators. The reason why the induced emission was discovered first in quantum physics is that we could not find a simple classical system that exhibits this effect. In classical electrodynamics, the amplified emission was found later, with the theory of the effect being quite involved.

The system of relativistic Landau levels is not equidistant; therefore, at first glance, an amplified synchrotron emission could be easily achieved provided an inverse population is formed in a system of relativistic electrons. However, there is another subtle point. 
By synchrotron, we typically mean emission at very high harmonics of the rotational frequency, such that the spectrum is continuous. In this case, waves emitted by an electron could be absorbed by electrons with a different energy. The inverse energy distribution can be formed only in a limited range of particle energies, say, at $E<E_0$, whereas at $E>E_0$, the level population decreases with the energy. Therefore, radiation from electrons with $E<E_0$ may be absorbed by electrons with $E>E_0$. For this reason, the synchrotron maser is possible only in two cases (see \citep{Melrose_book80} and references therein). The first is an extremely narrow energy distribution, which could hardly form in realistic conditions. In the second case, the inversely populated electrons radiate in the frequency range, where the dispersion of the waves is modified by the plasma, 
$(\omega/ck)-1\ge\gamma^{-2}$, where $\gamma$ is the characteristic Lorentz factor of electrons. Then, both emission and absorption are suppressed by the Razin effect, with the suppression being stronger for higher energy electrons. In ring-like distributions, the inverse population is formed at smaller energies; therefore, the induced emission of these electrons is less suppressed than the absorption by higher energy electrons. Then, the total absorption coefficient becomes negative, which implies maser emission.

It was noticed \citep{Sazonov70} that the dispersion of electromagnetic waves may be sufficiently modified by the relativistic electrons themselves provided their density is large enough. 
For this case, the theory of the synchrotron maser instability has been developed both  for spherically symmetric \citep{SagivWaxman02,Gruzinov_Waxman19} and for ring-like electron distributions \citep{Lyubarsky06}. The growth rate of the maser instability is estimated as
 \begin{equation}
 \kappa\sim 0.1 \frac{\Omega_B^{5/4}}{\Omega_p^{1/4}},
 \end{equation}
 where
 \begin{equation}
 \Omega_B=\frac{eB}{m_ec\gamma};
 \qquad\Omega_p=\sqrt{\frac{4\pi e^2 N}{m_e\gamma}}
 \label{Omega_B,Omega_p}\end{equation}
 are the relativistic Larmor and plasma frequencies, respectively. 
 The characteristic frequency of the amplified waves is:
 \begin{equation}
 \omega\sim \frac{\Omega_p^{5/4}}{\Omega_B^{1/4}}.
 \end{equation}
The above result was obtained under the condition $\Omega_B<\Omega_p$. In relativistic sources, we can conveniently use the magnetization parameter, defined as twice the ratio of the magnetic to the plasma energy density in the source:
\begin{equation}
 \sigma=\frac{B^2}{4\pi m_ec^2\gamma N}=\left(\frac{\Omega_B}{\Omega_p}\right)^2.
\end{equation}
Then, the characteristic frequency of amplified waves and the growth rate may be respectively presented as:
\begin{equation}
 \kappa\sim 0.1\sigma^{1/4}\Omega_B; \qquad\omega\sim \sigma^{-1/4}\Omega_p;\qquad \sigma<1.
\label{maser_sigma<1}\end{equation}
Because of a very weak dependence on $\sigma$, the emission frequency of the maser does not much exceed the plasma frequency. 

Above, only radiation at high harmonics of the rotational frequency, $\Omega_B$, was discussed. Note that the few first harmonics do not overlap if the electron energy distribution is not too wide, $\Delta E<E$. Therefore, the maser emission at the first harmonic is possible, even in a vacuum. In a strongly magnetized plasma, $\sigma>1$, the Larmor frequency exceeds the plasma frequency; therefore, waves with frequencies of the order of $\Omega_B$ could propagate. Therefore, in this case, the maser emits at the main and the few first harmonics, with the instability growth rate being (e.g., \citep{Aleksandrov_eyal84}):
\begin{equation}
 \kappa\sim 0.3\sigma^{-1/3}\Omega_B\qquad \omega\sim \Omega_B=\sigma^{1/2}\Omega_p;\qquad \sigma>1.
\label{maser_sigma>1}\end{equation}
One sees that the maser action is possible at any magnetization, provided an inverse electron population is formed. As, in the systems of interest, the magnetization is not too low and not too big, the characteristic emission frequency may be generally presented as: 
\begin{equation}
 \omega\sim \zeta\Omega_p,
\label{maser_frequency}
\end{equation}
 where $\zeta\sim$ a few. 

\subsection{Coherent Emission from the Front of a Relativistic, Magnetized Shock}

The reverse level population means that in some energy range, the particle energy distribution function grows faster with the energy than the thermal distribution.
This distribution could be formed only in highly non-equilibrium systems under rather non-trivial conditions. 

A ring-like particle distribution is naturally formed at a front of the collisionless shock in a magnetized flow because this shock is mediated by the Larmor rotation: when the upstream flow enters the shock, the bulk velocity sharply drops, and particles begin to rotate in the enhanced magnetic field. It was assumed \citep{Langdon_etal88} that at relativistic shocks, synchrotron maser instability develops. Numerical simulations of relativistic, magnetized shocks \citep{Hoshino92,Gallant92,Sironi_Spitkovsky11,Iwamoto_etal17,Iwamoto_etal18,Iwamoto_etal19,Plotnikov_Sironi19,Babul_Sironi20} revealed strong electromagnetic precursors with the characteristic frequencies compatible with the estimate (\ref{maser_frequency}).

Note that the frequencies (\ref{maser_frequency}) are in fact given in the downstream frame. In the upstream frame, they are Lorentz-boosted so that the frequencies of the emitted waves are well above the local plasma and Larmor frequencies. Therefore, the precursor waves propagate away relatively easily. 

The simulations show that a ring-like distribution is formed at the front of the shock, provided the magnetization of the flow is high enough, $\sigma\ge 10^{-3}$. However, we cannot certainly attribute the precursor emission to the maser mechanism. The ring is immediately destroyed just beyond the front, so that the width of the unstable region in the shock frame is only $\sim 2c\Omega_B^{-1}$. As the growth rate of the instability is less than $\Omega_B$, see Eqs.(\ref{maser_sigma<1}) and (\ref{maser_sigma>1}), a seed wave could not be amplified considerably while crossing the shock. The waves propagating within the shock plane perpendicular to the shock normal could be amplified, but then the precursor should be attributed to the scattering of the amplified waves. It is possible that the intrinsic unsteadiness of the shock front leads to strong fluctuations in the charge density so that the thus-formed charged bunches rotate in the magnetic field and emit coherent radiation. In any case, the emitted frequency is of the order of a few $\Omega_B$ at $\sigma>1$ and a few $\Omega_p$ at $\sigma<1$, which is compatible with our expectation from the synchrotron maser emission. Despite this uncertainty, I refer to this emission as the synchrotron maser emission, because this term is widely used in the literature. 

In the referenced simulations, the authors found that the emitted energy reaches a few percent of the inflow power at $\sigma\sim 0.1$ and decreases toward both large and small magnetizations.
The strength parameter of the electromagnetic waves is defined as the ratio of the cyclotron frequency in the wave field to the wave frequency:
\begin{equation}
 a=\frac{eE}{m_ec\omega},
\end{equation}
where $E=B$ is the amplitude of the wave. The cases $a\ll 1$ and $a\gg 1$ correspond, respectively, to non-relativistic and highly relativistic oscillations of electrons in the wave. This quantity is a relativistic invariant. Let us define the efficiency of the maser emission, $\chi$, as the ratio of the emitted power to the total power of the inflow, both measured in the shock front reference frame. Using Eq. (\ref{maser_frequency}), we find that if the inflow Lorentz factor is $\gamma$, the strength parameter of the emitted waves is \citep{Iwamoto_etal17,Plotnikov_Sironi19}:
\begin{equation}
 a\sim\frac{\chi^{1/2}}{\zeta}\gamma.
\label{strength}\end{equation}
One sees that a highly relativistic shock emits strong electromagnetic waves, which could have profound influence on their interaction with the surroundings.

The polarization of the radiation is an important issue. Most of the available simulations of the maser emission from relativistic shocks were performed either in 1D or 2D with the simulation plane perpendicular to the background magnetic field. No information on polarization could be inferred from these simulations. In \citep{Gallant92}, 2D simulations were performed for the magnetic field in the simulation plane; it was reported that the outgoing radiation is 100\% polarized perpendicular to the magnetic field. Conversely, both polarization modes were found in other studies \citep{Iwamoto_etal18,Plotnikov_Sironi19a}, where both 2D and 3D simulations were performed. The last result seems to be more compatible with theoretical expectations. The synchrotron emission is completely polarized only in the plane of the rotating particles, but the out-of-plane emission has an electric field component parallel to the magnetic field. Therefore, the total emission is polarized predominantly perpendicular to the magnetic field, but the degree of polarization never reaches 100\%.

The dependence of the synchrotron maser emission on the pre-shock temperature was numerically investigated \citep{Babul_Sironi20} for the strongly magnetized case, $\sigma\ge 1$. It was found that the maser efficiency is nearly independent of the temperature at $k_BT<0.03m_ec^2$, but sharply drops at higher temperatures, becoming two orders of magnitude smaller at $k_BT=0.1m_ec^2$. The reason is that even at such a relatively small energy spread of the upcoming electrons, the ring at the shock front becomes thick, $\Delta E\sim E$. As it was discussed in the previous subsection, the system emits at the first harmonics of the Larmor frequency at $\sigma>1$, only if these harmonics do not overlap, so that if $\Delta E< E$, which explains the above result. How weakly magnetized shocks depend on the upstream temperature remains unclear.

All the above refer to shocks in electron-positron plasma. The case of electron-ion plasma is more complicated. In electron-ion flow, the electrons transfer only a small fraction of the energy. However, the electrons lag behind ions in the field of intense precursor waves. Then, a longitudinal large-scale electric field is excited so that the electrons are accelerated, taking the energy from the ion flow \citep{Lyubarsky06}. Moreover the Raman scattering of the precursor waves produces plasma waves; this process leads to acceleration of the electrons and even to the formation of non-thermal electron distributions \citep{Hoshino08}. Therefore, the electrons could eventually reach an energy equipartition with the ions before the flow reaches the shock. In this case, the electron mass should be substituted by one-half the ion mass in Eq. (\ref{maser_frequency}) for the frequency of the emitted radiation so that the frequency of the precursor decreases. 
Numerical simulations of relativistic shocks in magnetized electron-ion flows \citep{Lyubarsky06,Hoshino08,Iwamoto_etal18} reveal a large-scale longitudinal electric field and the electron acceleration until the equipartition upstream of the shock. The tendency of the emitted frequency to decrease was also found \citep{Hoshino08}, but the relatively small ion-to-electron mass ratio in simulations (only 50) does not permit to firm conclusions. 

The outlined energy equilibration mechanism assumes that the strong-enough precursor is produced from the beginning, when the electrons are still light. 
However, in this case, the shock is mediated by the Larmor rotation of ions; it is not evident that the ring-like distribution of light electrons is formed. The synchrotron radiation of ions could be suppressed at this stage because the frequency is well below the electron plasma frequency; then, the equilibration mechanism is not involved. However, if the electron magnetization is not small, such that the electron Larmor frequency exceeds the radiation frequency, the low-frequency radiation becomes possible. This question has not been properly investigated theoretically. In numerical simulations, the shock is initiated by hitting the flow upon a rigid conductive wall. In this case, a strong electromagnetic perturbation is initially formed, which is sufficient to trigger the acceleration process. It is unclear how the process is triggered in reality.

Another problem is that the accelerated electrons have a wide quasithermal energy distribution, in which case the synchrotron maser emission may be suppressed. It was speculated \citep{Iwamoto_etal18} that after the precursor is quenched, cold undisturbed electrons enter the shock once again and the whole positive feedback cycle is initiated. More simulations are necessary to clarify the issue.

\subsection{Radiation from Reconnecting Current Sheets}

In this section, a sort of antenna mechanism is described. An antenna is a linear conductor carrying a variable current. Variable linear currents are naturally formed in the course of magnetic reconnection in a current
sheet, separating oppositely directed magnetic fields. In the reconnection process, the current
sheet breaks into a system of linear currents, which are called magnetic islands because of their
island-like appearance in 2D simulations. The process is highly variable so that we can easily imagine that variable currents are sources of electromagnetic waves. This analogy is not very useful because antennas emit in a vacuum, whereas the reconnection process occurs in plasma, the characteristic variability times being larger than the microscopic plasma times. The frequency of the emitted waves is lower than both the plasma and the Larmor frequency; therefore, the system should be described in terms of magnetic hydrodynamics (MHD).

Inasmuch as the parallel currents are attracted one to another, the islands continuously merge (see, e.g., \citep{Kagan_etal15}).  The merging
of two islands perturbs the magnetic field in the vicinity of the merging point, thus exciting
MHD waves around the reconnecting current sheet. There are generally three types of MHD waves: the Alfv'en wave and two magnetosonic
waves. In the simplest case
of cold plasma, only the Alfv'en and the fast magnetosonic (fms) waves remain. In the Alfv'en waves, the magnetic field lines oscillate due to the magnetic tension as stretched strings. These waves propagate along the magnetic field lines;
therefore, they do not transfer the energy away from the current sheet. Fast magnetosonic waves propagate across the magnetic field lines. These waves resemble sound waves, but instead of the gas pressure, the restoring force is provided by the magnetic pressure. In cold plasma, the fms velocity is (e.g., \citep{Appl_Camenzind88}):
 \begin{equation}
 v_{\rm fms}=\sqrt{\frac{B^2}{4\pi\rho c^2+B^2}}=\sqrt{\frac{\sigma}{1+\sigma}},
 \label{fms_velocity}\end{equation}
where $\rho$ is the plasma density; all quantities are measured in the plasma frame. The fms velocity is independent of the propagation direction; therefore, any merging event produces
a quasi-spherical fms pulse that propagates away, with the duration of the pulse being $\sim a/c$, where $a$ is the transverse size of the island. The size of magnetic islands scales with the Larmor radius of particles in the current sheet, $a\sim{\rm few}\,c/\Omega_B$.
Therefore, the reconnection process produces an emission with wavelength of the order of a dozen of Larmor radii of particles within the sheet. The fms pulses are clearly observed in simulations of the reconnection in the current sheet \citep{Philippov19}; it was found that they take away roughly 0.5\% of the magnetic energy dissipated in the course of the reconnection process.

In the fms wave, the plasma density oscillates along the propagation direction, just as in sound waves. The magnetic field is frozen into the plasma; therefore, plasma compression and rarefaction are accompanied by oscillations in the field strength. The electric field vanishes locally in the plasma frame; therefore, in the global frame where the plasma is on average at rest, the electric field satisfies the relation $\mathbf{E}+\frac 1c\mathbf{v\times B}=0$, where $v$ is the oscillation velocity. Thus, the electric field of the wave is perpendicular both to the background magnetic field and to the propagation direction. The fms wave is longitudinal as the hydrodynamic wave but it is a transverse electromagnetic wave. 

In the magnetically dominated plasma, $\sigma\gg 1$, the fms velocity is close to the speed of light, and when the wave propagates toward smaller plasma densities, such that $\sigma\to\infty$, it becomes a vacuum electromagnetic wave polarized perpendicular to the background magnetic field. This could be easily understood as follows: Consider a
vacuum electromagnetic wave superimposed on a homogeneous
magnetic field such that the electric field of the wave is
perpendicular to the background field. In this system, a seed charged particle oscillates due to drift in the crossed electric field of the wave and the background magnetic field. 
The velocity of the electric drift is independent of the particle charge; therefore, if a small amount of plasma is added to
the system, no electric current appears in the system and the
wave propagates as in vacuum.

The coherent radio emission could be produced in the course of reconnection in magnetically dominated plasma, provided the Larmor radius of particles in the current sheet is comparable with the radio wavelength. This mechanism was recently proposed as a source of radio emission from the Crab and Crab-like
pulsars \citep{Uzdensky_Spitkovsky14,Lyubarsky19,Philippov19}. In ref.\ \citep{Lyubarsky20}, the mechanism was applied to FRBs. 

\section{FRBs from Magnetar Flares}

In the previous section, mechanisms of coherent radio emission were outlined, which look promising in the FRB context. They all imply highly magnetized and relativistic plasma, which could be found in many potential FRB progenitors. Magnetars look to be the most promising sources of FRBs.  Initially, the FRB-magnetar connection was proposed on statistical ground \citep{PopovPostnov07} because the rate of magnetar flares is comfortably above the FRB rate, as distinct from, say, gamma-ray bursts.  The recent discovery of weak FRBs from the Galactic magnetar SGR 1935+2154 \citep{CHIME20,Bochenec20} lends strong support to magnetar models. In this section, I describe how the above-outlined coherent radiation mechanisms could work in magnetars.

\subsection{Magnetar Magnetosphere and Wind}

Magnetars are neutron stars with surface magnetic field $\sim 10^{15}$ G. Even a larger field, $\sim 10^{16}$ G, could be buried under the surface of the star. The notion of magnetars was introduced to astrophysics by Duncan and Thompson \citep{Duncan_Thompson92,Thompson_Duncan95}, who predicted that ultrastrong magnetic fields could be generated in proto-neutron stars, and demonstrated that the decay of this field could feed soft gamma repeaters (for a recent review of magnetars, see \citep{Kaspi_Beloborodov17}). Restructuring the magnetic field in a magnetar's magnetosphere produces X-ray flares with $10^{40}-10^{46}$ erg energies and with durations from a fraction of a second to a few minutes. 
 A rotating magnetar's magnetosphere emits a relativistic wind, similar to pulsar wind, which removes the rotational energy of the neutron star. Therefore, the neutron star spins down; the magnetar period increases to $\sim 1$ s for only $\sim 10$ years. 

In the magnetosphere of an active magnetar, the magnetic field lines are twisted; therefore, the field is non-potential. The twist currents in the magnetosphere are carried by electron--positron pairs produced by the current \citep{BeloborodovThompson07,Beloborodov13a}: when the density of the pairs drops below the limit sufficient to maintain the current, the induced electric field (the displacement current) accelerates particles and triggers an avalanche, producing new pairs. Nevertheless, the stability considerations imply that in the persistent state, the magnetic field could hardly strongly exceed the potential field; therefore, for rough estimates of the field strength, we could use the dipole field. 

The rotating magnetosphere becomes open at the light cylinder radius:
\begin{equation}
 R_L=cP/2\pi =4.8\times 10^9P\,\rm cm,
\label{RL}\end{equation} 
where $P$ is the rotational period. The magnetic field at the light cylinder is estimated as:
\begin{equation}
 B_L=\frac{\mu}{R_L^3}=9\cdot 10^3 \frac{\mu_{33}}{P^{3}}\,\rm
G,
 \label{BL}\end{equation}
where $\mu$ is the magnetic moment of the star.  In magnetars, the surface magnetic field is $B_*\sim 10^{15}$ G therefore the magnetic moment is $\mu\sim B_*R^3_*\sim 10^{33}$ G$\cdot$cm$^3$, where $R_*\approx 10^6$ cm is the neutron star's radius. The pairs are continuously ejected from the magnetosphere, forming a magnetar wind. The particle flux in the wind from a strongly twisted magnetar magnetosphere was estimated as \citep{Beloborodov20}:
\begin{equation}
 \dot{\cal N}\sim{\cal M}\frac{c\mu}{eR_LR_{\pm}}=2.5\times 10^{39}\frac{{\cal M}_3\mu_{33}^{2/3}}{P}\,\rm s^{-1},
\label{Ndot}\end{equation}
where ${\cal M}$ is the pair multiplicity  (should  not be confused with the multiplicity parameter used in the pulsar theory) and $R_{\pm}\sim 5\times 10^6\mu^{1/3}_{33}$ cm is the distance from the star where the magnetic field falls to $10^{13}$ G, such that the pair production
stops.  The maximal multiplicity was estimated to be of the order of ${\cal M}\sim 10^3$ \citep{Beloborodov20}, but it should be stressed that the process of pair production in magnetar's magnetospheres is highly uncertain therefore one cannot exclude that the pair loading of the magnetar wind is much larger or smaller than that given by eq.\ (\ref{Ndot}). 

The magnetar wind is strongly magnetized. 
Beyond the light cylinder, the field is wound up by the rotating magnetosphere so that in the far zone, the azimuthal field dominates:
\begin{equation}
 B_{\rm wind}=B_L\frac{R_L}{R}.
\end{equation}
The initial magnetization parameter of the magnetar wind, which is
defined as the ratio of the Poynting flux to the rest mass energy flux in the wind, is:
\begin{equation}
\eta=\frac{B_L^2R_L^2}{\dot{\cal N}m_ec}=2.5\cdot 10^4\frac{\mu_{33}^{4/3}}{{\cal M}_3P^3}.
 \label{eta}\end{equation}
The magnetization parameter is the maximal Lorentz factor achievable by the wind if the
magnetic energy is completely converted to kinetic energy.

Beyond the light cylinder, the strongly magnetized wind accelerates linearly with distance $\gamma\sim R/R_L$, until it reaches the fast
magnetosonic point, $\gamma\sim\eta^{1/3}$. If there is no dissipation, the wind
accelerates very slowly beyond the fast magnetosonic point, $\propto(\ln R)^{1/3}$ \citep{Beskin98}. Therefore, we can take the Lorentz factor of the wind in the far zone, $R\gg \eta^{1/3}R_L$, to be:
 \begin{equation}
\tilde{\gamma}_{\rm wind}=3\eta^{1/3}=90\frac{\mu_{33}^{4/9}}{{\cal M}^{1/3}_3P}.
 \end{equation}
The magnetic dissipation could lead to a gradual acceleration of the wind in the equatorial belt,
where the magnetic field changes sign every half of a period
\citep{Lyubarsky_Kirk01,Kirk_Skjeraasen03}; then, the Lorentz factor of the wind exceeds
$\tilde{\gamma}_{\rm
wind}$ and may even reach $\eta$.  The efficiency of the dissipation is quite uncertain therefore in all formulas below I retain $\gamma_{\rm wind}\ge\tilde{\gamma}_{\rm
wind}$ as a free parameter. The ratio of the Poynting to the plasma kinetic flux in the wind may be generally presented as:
\begin{equation}
\sigma_{\rm wind}=\frac{B^2}{4\pi mc^2\gamma_{\rm wind}N}=\frac{\eta}{\gamma_{\rm wind}}=
 280\frac{\mu_{33}^{8/9}}{{\cal M}^{2/3}_3P^2}\frac{\tilde{\gamma}_{\rm wind}}{\gamma_{\rm wind}}, \label{sigma_wind}\end{equation}
where $N$ is the pair density in the wind.

\subsection{Magnetar Flares and FRBs}

The energy of magnetar flares is sufficient to feed FRBs, even if a small fraction is converted to radio waves. The flares are produced by a rapid restructuring of the magnetar's magnetic field. The duration of FRBs is compatible with the Alfv'en crossing time of the inner magnetosphere, which is the characteristic time of the development of MHD instabilities, which trigger the flares. The duration of the flares significantly exceeds this time, which could be attributed to the relatively long time necessary to radiate the released energy away. It is also possible that during the flare, a few magnetic explosions happen (see, e.g., \citep{Yuan_etal20}). In this case, the question remains why only one explosion typically produces an FRB.

In the magnetar magnetosphere, both the Larmor and the plasma frequencies are far above the radio band. Therefore, if the source of FRBs was deep within the magnetosphere, it should have emitted MHD waves. However, these waves are unable to escape because of non-linear interactions, as will be shown in Section 4.5. Therefore, FRBs could be produced only in the outer magnetosphere or in the magnetar wind. Far from the neutron star, the magnetic energy density is not sufficient in order to produce a short, powerful burst. However, the necessary energy could be delivered to the far zone by a strong magnetic pulse excited during a sudden rearrangement of the magnetosphere, which gives rise to 
the magnetar flare. Let us consider the properties of such a pulse.

A rapid restructuring of the magnetosphere 
 produces a large-scale MHD perturbation that propagates
outward, sweeping the magnetic field lines into a 
pulse of length $l=c\tau=3\times 10^7\tau_{\rm ms}$, where $\tau$ is the duration of the magnetic restructuring process. The last is of the order of the Alfv'en crossing time of the inner magnetosphere, i.e., of a few to a few tens of stellar radii. The pulse opens the magnetosphere, and propagates further out into the magnetar wind. As discussed in Section 2.6, only fms waves can propagate across the magnetic field lines. Therefore, independently of the initial magnetic configuration,  we deal an fms pulse in the far zone. The amplitude of the pulse may be conveniently presented as:
\begin{equation}
 B_{\rm pulse}=\sqrt{\frac{L_{\rm pulse}}c}\frac 1R=3.8\times 10^8\frac{L_{\rm
pulse,47}^{1/2}}{P}\frac{R_L}R\, \rm G,
 \label{Bpulse}
\end{equation}
where $L_{\rm pulse}$ is the isotropic luminosity associated with the pulse. The total
energy in the pulse is:
\begin{equation}
 {\cal E}=L_{\rm pulse}\tau=10^{44}L_{\rm pulse,47}\tau_{\rm ms}\,\rm erg.
 \label{energy-total}
\end{equation}
In the far zone, the amplitude of the pulse significantly exceeds the background magnetic field. 

Just beyond the light cylinder, the pulse enters the magnetar wind. Neglecting plasma inertia, i.e., in the limit $\sigma\to\infty$, the pulse may be considered purely electromagnetic. It propagates nearly at the speed of light; the magnetic field is purely azimuthal, whereas the electric field is poloidal and equal to the magnetic field $E_{\rm pulse}=B_{\rm pulse}$. The plasma is squeezed in the pulse and pushed forward; the plasma velocity with respect to the wind is estimated as the velocity of the zero electric field frame, $v'=E'/B'=B'_{\rm pulse}/(B'_{\rm wind}+B'_{\rm pulse})$, where the prime refers to quantities in the wind frame.
The corresponding Lorentz factor is:
\begin{equation}
 \Gamma'=\sqrt{\frac{B'_{\rm pulse}}{2B'_{\rm wind}}}=\frac 12\sqrt{\frac{B_{\rm pulse}}{B_{\rm
wind}}}=100\frac{L_{{\rm pulse,}47}^{1/4}P}{\mu_{33}^{1/2}}.
 \label{Gamma}\end{equation}
 In the lab frame, the Lorentz factor of the pulse is $\Gamma=2\Gamma'\gamma_{\rm wind}$.
 
 The pulse itself, i.e.,
the waveform, moves with the fms velocity (\ref{fms_velocity}) with respect to the local plasma velocity. For a large magnetization $\sigma\gg 1$ and a high amplitude of the pulse $B_{\rm pulse}\gg B_{\rm wind}$, the difference between the velocity of the waveform and the speed of light is estimated in the wind frame as:
\begin{equation}
 \frac{c-v'_{\rm form}}c\sim (\sigma\Gamma'^2)^{-1}\sim\frac{B_{\rm wind}}{\sigma\gamma^2_{\rm wind}B_{\rm pulse}},
\label{delta_v}\end{equation}
so we can safely assume that the pulse moves with the speed of light. The pulse moves through the magnetar wind as a
propagating wave so that the plasma enters the pulse through the front part and eventually leaves it through the rear part. Within the pulse, the plasma moves with respect to the wind with the large Lorentz factor (\ref{Gamma}). Therefore, the plasma is dragged within the pulse to a large distance and is very slowly substituted by the wind plasma.

The dependence of the waveform velocity on the local density and magnetic field, which vary
across the pulse, could lead to the non-linear steepening of the pulse. However, this velocity is very close to the speed of light (see Eq. (\ref{delta_v})); therefore, the nonlinearity is weak even if the amplitude of the pulse is very large \citep{Levinson_vanPutten97,Lyubarsky03,Lyutikov10}. The physical reason for this is that the displacement current significantly exceeds the conductivity current, so the pulse propagates nearly as if in a vacuum. Moreover, the waveform propagates with the fms velocity in the local plasma frame, which moves at relativistic speeds if the pulse amplitude exceeds the background field, see Eq.(\ref{Gamma}). Substituting the estimated above parameters of the wind and pulse in Eq. (\ref{delta_v}), the non-linear steepening scale is estimated as
\citep{Lyubarsky20}:
\begin{equation}
 R_{\rm steep}=\frac{cl}{c-v_{\rm form}}=8\Gamma^2\sigma_{\rm pulse}l=10^{13}\sigma_{\rm pulse}\gamma^2_{\rm wind}\frac{L_{\rm pulse, 47}^{1/2}P^2\tau_{\rm ms}}{\mu_{33}}\,\rm cm.
 \label{steepening1}\end{equation}

The above considerations implicitly assume that the mean
field in the wind occurs in the same direction as the initial
field of the pulse. However, we can imagine a situation
where the field in the wind is opposite to the field in the pulse. In an ideal case, this does not affect the obtained results because we can imagine a pulse within which the magnetic field changes sign such that a current sheet separates the domains of the opposite polarity. The electromagnetic stress is a quadratic function of the fields; therefore, a solution of ideal
MHD equations is not affected if we reverse fields in any bundle of the magnetic field lines and insert an appropriate
current sheet, provided that the plasma in the sheet is light
enough (so that the overall inertia is not affected). In reality, the current sheet is unstable so that the new flux could annihilate with the flux in the pulse. The flux in the pulse, $B_{\rm pulse}l$, significantly exceeds the flux in the wind, $B_{\rm wind}R=B_LR_L$; therefore, the pulse will not be destroyed.
However, the field annihilation heats the plasma and 
the heated plasma expands, transforming heat into kinetic energy.
A detailed analysis of this process has not yet been performed. 

\subsection{FRBs Produced by Relativistic Shocks from Magnetar Flares}

As shown in Section 2.5, relativistic magnetized shocks are efficient sources of coherent emission. Let us discuss how such shocks could be produced by outflows from magnetar flares and whether the properties of their emission are compatible with the observed properties of FRBs. 

It was proposed \citep{Lyubarsky14} that an FRB is produced when the electromagnetic perturbation from the flare (magnetic pulse described in Section 3.2) reaches the nebula inflated by the magnetar wind in the surrounding gas. The inner boundary of the nebula is determined by the balance of the pressure within the nebula and the dynamic pressure of the magnetar wind:
\begin{equation}
 R_s=\frac{B_LR_L}{\sqrt{4\pi p}}=1.2\times 10^{15}\frac{\mu_{33}}{Pp_{-4}^{1/2}}\,\rm cm,
\end{equation}
where $p$ is the pressure within the nebula. The last quantity is highly uncertain; normalization by $10^{-4}\,\rm dyne/cm^2$ was chosen because this value is roughly compatible with the observational data for the nebula surrounding FRB 121102 \citep{Beloborodov17}.

When the electromagnetic pulse arrives at the wind termination
shock, it pushes the plasma outward like a magnetic piston. A forward shock propagates through the magnetized and relativistically hot
plasma of the nebula, whereas a reverse shock enters the magnetic piston. Between the shocks, a contact discontinuity separates
the shocked plasma of the nebula from the magnetic piston. At the
contact discontinuity, the magnetic pressure of the pulse is balanced
by the bulk pressure of the relativistically hot plasma entering the forward shock. The pressure balance condition yields the Lorentz factor of the contact discontinuity:
\begin{equation}
 \Gamma_{\rm cd}=\left(\frac{B_{\rm pulse}^2}{32\pi p}\right)^{1/4}=150\frac{L_{\rm pulse,47}^{1/4}P}{\mu^{1/2}_{33}}.
\end{equation}
The plasma in the nebula is mildly magnetized, $\sigma\le 1$; therefore, the forward shock moves relative to the contact discontinuity only mildly relativistically; we can roughly take the Lorentz factor of the shock to be $\Gamma_{\rm cd}$. 

The frequency of the maser radiation in the shock frame is estimated by Eq. (\ref{maser_frequency}). The plasma pressure beyond the forward shock is of the order of the magnetic pressure in the pulse. Therefore, the relativistic plasma frequency (\ref{Omega_B,Omega_p}) at the shock front is roughly equal to the Larmor frequency of particles rotating with the Lorentz factor $\Gamma_{\rm cd}$ in the field $B'_{\rm pulse}=B_{\rm pulse}/\Gamma_{\rm cd}$. Making the Lorentz transform, we estimate the observed frequency as:
\begin{equation}
 \nu\sim\frac{\zeta eB_{\rm pulse}}{2\pi m_ec\Gamma_{\rm cd}}=28\zeta\frac{ L_{\rm pulse,47}^{1/4}p_{-4}^{1/2}P}{\mu_{33}}\,\rm MHz.
\end{equation}
The observed duration of the pulse is determined by the time spread of the arrival of radiation emitted by the fraction of the forward shock, subtending the angle $\Gamma^{-1}_{\rm cd}$. This yields:
\begin{equation}
 \tau_{\rm obs}\sim\frac{R_s}{2c\Gamma_{\rm cd}^2}=0.9\frac{\mu_{33}}{L^{1/2}_{\rm pulse,47}P^4p_{-4}^{1/2}}\,\rm s.
\end{equation}
The observed duration of the burst, $\tau\sim 1$ ms, and the frequency, $\nu\sim 1$ GHz, may be reproduced only by assuming a long-period magnetar, $P\sim 5$ s, and the power, $L_{\rm pulse}\sim 10^{38}$ erg/s. However, the nebula around such a long-period magnetar could hardly have large energy density and pressure, whereas decreasing $p$ demands a further increase in the burst power. In the original paper \citep{Lyubarsky14}, FRBs were attributed to very rare superflares with the total energy of $\sim 10^{48}$ erg, which could occur when the neutron star eventually becomes unstable to a dynamic overturning instability that destroys most of its dipole moment in a single event \citep{Eichler02}. Later-discovered repeating FRBs, as well as the ubiquity of single FRBs, indicated that the energy beyond a typical FRB could hardly ever be that large. Moreover, the discussion in Section 2.5 of simulations casts doubt of the possibility of maser emission from shocks in a relativistically hot medium.

To resolve these difficulties, it was assumed that the shock arises in the magnetar wind at distances of the order of $10^{14}$ cm \citep{Beloborodov17,Beloborodov20,Yuan_etal20}. In this case, the pulse acts as a piston driving a blast wave into the cold pre-explosion wind. The plasma in the pulse moves with the Lorentz factor (\ref{Gamma}) with respect to the wind. In a highly magnetized medium, the shock propagates with respect to the downstream plasma with the fast magnetosonic Lorentz factor $\sqrt{\sigma}$, Eq. (\ref{fms_velocity}). Therefore, in the lab frame, the shock Lorentz factor is $\Gamma_{\rm shock}=2\Gamma\sigma_{\rm wind}^{1/2}$. In the downstream frame, the particles rotate with the Lorentz factor $\Gamma'$ in the magnetic field $B'_{\rm pulse}=B_{\rm pulse}/\Gamma$. The frequency of the observed maser emission is obtained by the Lorentz transform:
\begin{equation}
 \nu=\zeta\frac{eB'_{\rm pulse}}{2\pi m_ec\Gamma'}\Gamma
=0.5\zeta\frac{\mu_{33}^{1/2}L^{1/2}_{\rm pulse, 47}}{PR_{14}}\,\rm GHz,
\label{frequency1}\end{equation}
which is compatible with observations (recall that $\zeta\sim$ a few, see Eq. \ref{maser_frequency}), if the radiation occurs at the distance $R\sim 10^{14}$ cm.
The observed duration of the burst is:
\begin{equation}
 \tau_{\rm obs}=\frac{R}{2c\Gamma^2_{\rm shock}}=0.01\frac{\mu_{33}R_{14}}{\sigma_{\rm wind}\gamma^2_{\rm wind}L^{1/2}_{\rm pulse,47}P^2}\, \rm s.
\label{tau_obs}\end{equation}
The observed duration, $\sim 1$ ms, may be obtained only if the magnetar wind is mildly relativistic,  $\gamma_{\rm wind}\sim$ a few, and moderately magnetized, $\sigma_{\rm wind}\sim 1$. Moderate Lorentz factor and magnetization of the wind are also required by the condition that the deceleration scale of the piston, $\sim \Gamma^2l$, does not significantly exceed $\sim 10^{14}$ cm because in the opposite case, too little energy is emitted in the GHz band. 

This model is based on the assumption that the 
pulse may be considered as a piston, i.e., the front part of the pulse is a flat wall pushing the plasma of the wind outward. However, initially, the pulse has a smooth shape, with the magnetic field gradually growing to the maximum at the scale of $\sim l=c\tau$.
 The shock could only arise due to the non-linear steepening 
discussed in Section 3.2. The larger the local magnetic field, the closer the velocity of the waveform to the speed of light, see Eq. (\ref{delta_v}). Fractions of the pulse with a larger field overtake the foot of the pulse, where the magnetic field approaches the background field; therefore, initially, a weak shock arises at the foot of the pulse. The field jump at the shock grows gradually as the fractions of the pulse with a larger field reach the shock. The shock velocity increases when the jump increases; therefore, the steepening process slows. The shock matches the amplitude of the pulse when the point of the maximum field arrives at the shock. This occurs at the distance given by Eq. (\ref{steepening1}). Note that the magnetization $\sigma_{\rm pulse}$ in this formula is determined by the magnetization just downstream of the shock because these parts of the pulse overtake the shock. There, the matter is already substituted by the matter of the wind; therefore, we can substitute $\sigma_{\rm pulse}=\sigma_{\rm wind}$ into this expression. 
Substituting $R_{\rm steep}$ into Eq. (\ref{frequency1}), we obtain:
\begin{equation}
 \nu=5\frac{\zeta}{\sigma_{\rm wind}\gamma_{\rm wind}^2}\frac{\mu^{3/2}_{33}}{L^{1/4}_{\rm pulse, 47}P^3\tau_{\rm ms}}\,\rm GHz.
\label{frq}\end{equation}
We again see that we obtain a reasonable observed frequency if  $\gamma_{\rm wind}\sim$ a few and $\sigma_{\rm wind}\sim 1$. 
Substituting $R_{\rm steep}$ into Eq. (\ref{tau_obs}), we find that $\tau_{\rm obs}=\tau\sim 1$ ms, independent of other parameters of the pulse. 

At the distance (\ref{steepening1}), the shock passes across a significant fraction of the pulse, thus releasing a significant fraction of the pulse energy. Therefore, the isotropic energy of the FRB is only proportional to the isotropic energy of the pulse:
\begin{equation}
 {\cal E}_{\rm FRB}=\chi {\cal E}=10^{41}\chi_{-3}L_{\rm pulse,47}\tau_{\rm ms}\,\rm erg.
\end{equation}
The model easily explains even bright FRBs. As the radiation frequency very weakly depends on the pulse energy, whereas the burst duration does not depend on it at all, the model can explain the large scatter of the observed bursts in energy. For example, taking small pulse energy and luminosity, say, $L_{\rm pulse}\sim 10^{42}-10^{41}$ erg/s, and the magnetar period of a few seconds, we obtain the same radiation frequency of about GHz, whereas the energy of the burst is only $\sim 10^{35}$ erg, which is compatible with the data for an FRB from galactic magnetar SGR 1935+2154.  Some repeaters exhibit bursts over a broad range of frequencies from ~400 MHz to 8 GHz. This could be attributed to the variations  in $\sigma_{\rm wind}$ and $\gamma_{\rm wind}$. 

The model of the shock in the magnetar wind successfully confronted  
to observations, provided the magnetar wind is only mildly relativistic and moderately magnetized. This implies that the initial magnetization of the wind, $\eta$, is of the order of a few so that the wind is heavily loaded by the plasma. According to Eq. (\ref{eta}), the required plasma multiplicity is ${\cal M}\sim 10^7$, which is about four orders of magnitude larger than the available magnetar models could provide. It was claimed \citep{Beloborodov17,Beloborodov20} that in a highly twisted pre-flare magnetosphere, the effective magnetic moment increases; then, the observed parameters of the burst may be achieved at $\gamma_{\rm wind}\sim 10$ and  $\sigma_{\rm wind}\sim$ a few, so that $\eta$ may reach a few dozen. Still, this demands an extremely large multiplicity. What mechanism could load the wind by the plasma so efficiently is still unclear. 

It was assumed \citep{Yuan_etal20} that during the magnetar flare, repeating magnetic explosions eject a chain of plasmoids. The magnetic field lines gradually reconnect behind each plasmoid, generating a wind far stronger than the normal spindown. Repeating ejections drive blast waves in the amplified wind, producing FRBs. This model successfully reproduces the observed properties of an FRB from the Galactic magnetar SGR 1935+2154 \citep{CHIME20,Bochenec20}, and specifically two subburst components separated by 30 ms. However, the model is based on axisymmetric simulations. In 3D, these plasmoids become unstable and dissipate for about 10 Alfven times \citep{Riddhi_etal20}. Therefore, they could hardly escape the magnetosphere. However, copious pair production in the course of the magnetar flare could load the wind, at least temporarily. The model deserves further development.

To resolve the problem of a highly mass loaded magnetar wind, it was suggested that the shock arises when the pulse collides with the mildly relativistic baryon cloud ejected by a previous flare \citep{Beloborodov17,Metzger19,Beloborodov20,Margalit20a,Margalit20b,Xiao_Dai20}. It is an observational fact that a significant amount of baryonic matter is released in giant magnetar flares \citep{Gaensler05,Gelfand_etal05,Granot_etal06}. The model assumes that not earlier than 1 day before the FRB, a strong flare ejected the required amount of material. Then, the magnetic pulse drives a shock in the cloud. However, the magnetization rapidly decreases within an expanded cloud, which means that the conditions for the synchrotron maser could be violated in this shock. Even if the magnetization of the cloud is not too small, the maser emission from a shock in the baryonic plasma is determined by proton plasma and cyclotron frequencies, which are very low. Therefore, the observed FRBs could hardly be produced by shocks in such clouds. After the cloud is ejected, the magnetar wind flows around it and forms a bow shock from behind. Therefore, before the pulse enters the cloud, it passes through a magnetized electron-positron plasma of the shocked magnetar wind. The pulse drives the shock in this plasma; therefore, we can speculate that the emission is produced by this shock. However, this plasma is relativistically hot; therefore, the synchrotron maser may be disabled in this case, see \citep{Babul_Sironi20} and the discussion in Section 3.3. Therefore, for a while, there is no satisfactory solution to the problem of a mildly relativistic, heavily-plasma-loaded medium required for the synchrotron maser model. 

In all the above models, the shock propagates outward; therefore, the emission frequency, which scales with the local Larmor frequency, decreases. So, the sometimes-observed 
downward frequency drift in FRBs \citep{CHIME19,Hessels19} could be naturally explained by the synchrotron maser models \citep{Metzger19,Beloborodov20,Margalit20a}.

The synchrotron maser emission is polarized perpendicular to the background magnetic field. This agrees with the typically observed high linear polarization. However, this implies that in repeaters, the polarization position angle remains the same, namely, along the rotation axis of the magnetar. This is incompatible with the diverse polarization angle swings observed in FRB 180301 \citep{Luo_etal20}. As discussed in Section 2.5, the synchrotron radiation could not be fully polarized. Therefore, the observed 100\% degree of polarization in a few FRBs \citep{Gijjar18,Michili18,oslowski19} could hardly be explained within the scope of the synchrotron maser model.

\subsection{FRBs from Magnetic Reconnection in the Upper Magnetar Magnetosphere}

As discussed in Section 2.6, in the course of magnetic reconnection, fast magnetosonic (fms) waves are efficiently generated, and they are converted to electromagnetic waves when propagating toward a decreasing plasma density. The wavelength of this radiation is of the order of the size of magnetic islands in the current sheet, which, in turn, are scaled with the Larmor radius of the particles in the sheet. The violent reconnection occurs during the magnetar flare when the unstable magnetic configuration is disrupted 
(e.g., \citep{Parfrey_etal13,Carrasco_etal19}). However, all microscopic plasma parameters are very small in the inner magnetosphere, so it is unclear why the waves with wavelengths $\sim 10$ cm
could be generated with a sufficiently large power. In any case, Section 4.5 shows that these waves could not escape because of non-linear interactions.

FRBs could be produced when the magnetic perturbation from the magnetar flare (magnetic pulse) reaches the current sheet separating, just beyond the light cylinder, the oppositely directed magnetic fields \citep{Lyubarsky20}. When the magnetic pulse arrives at the sheet, the sharp acceleration and compression cause violent reconnection. We can speculate that the current sheet is destroyed by the
Kruskal–Schwarzschild instability \citep{Lyubarsky10,Gill_etal18}, which is the magnetic counterpart of the Rayleigh–Taylor instability, so that the field line tubes with the oppositely-directed fields fall into the magnetic pulse, forming multiple small current sheets scattered over the body of the pulse. Within each of the small current sheets, the reconnection process occurs via formation and merging of magnetic islands, which procudes an fms noise. This noise is converted, as was outlined in Section 2.6, into radio emission. Since the sources of the noise are distributed in the body of the magnetic pulse but not concentrated at the front part, the duration of the
observed radiation burst is of the order of the duration of the pulse, $\tau_{\rm obs}\sim l/c=\tau$. 

Assuming the above picture, let us estimate the parameters of the outgoing radiation.
The magnetic pulse from the flare enters the magnetar wind just beyond the light cylinder. The plasma is squeezed in the pulse and pushed forward with the Lorentz factor (\ref{Gamma}). In the magnetar wind, a current sheet separates two magnetic hemispheres, with the shape of the sheet having been likened to
a ballerina's skirt.
The total reconnecting magnetic flux
may be estimated as $B_LR_L^2$. Within the pulse, the stripe with the
oppositely-directed fields is compressed $B_{\rm pulse}/B_L$ times. Then, the total energy of annihilated fields is roughly:
\eqb
{\cal E}\sim\left(B_{\rm pulse}/B_{\rm wind}\right)B^2_LR_L^3=\left(L_{\rm pulse}/c\right)^{1/2}B_LR_L^2.
\eqe
According to the simulations \citep{Philippov19}, the fraction $f\sim 0.005$ of the reconnecting
magnetic energy is emitted in the form of fms waves. Making use of Eqs. (\ref{RL}) and (\ref{BL}), we estimate the isotropic energy of the radio burst as:
\eqb
{\cal E}_{\rm FRB}=f{\cal E}=3.8\cdot 10^{39}\frac{f_{-2}\mu_{33}L_{{\rm
pulse,}47}^{1/2}}{P}\,\rm erg.
 \label{energy}\eqe
 
The wavelength of the emitted waves is of the order of the size of magnetic islands within the current sheet, which are 10--100 times larger than the width of the current sheet
\citep{Philippov19}. Therefore, we can estimate the emitted frequency in the comoving frame as:
 \eqb
\omega'=\frac c{\xi a'},
\label{freq_reconn}\eqe
where $a'$ is the width of the sheet in the comoving frame, $\xi\sim 10-100$. 
 The reconnection occurs via the collisionless tearing
instability; therefore, the width of the sheet is of the order of a few Larmor radii:
\eqb
a'=\zeta\frac{\varepsilon_T}{eB'_{\rm pulse}},
\eqe
where $\varepsilon_T$ is the characteristic thermal energy of
the pairs within the sheet,  $\zeta\sim$ a few. To estimate $\varepsilon_T$, let us consider the energy and the pressure balance within the sheet.

Within the sheet, the pressure of the external magnetic
field is balanced by the pressure of hot pairs:
\eqb
\frac 13N'\varepsilon_T=\frac{B'^2_{\rm pulse}}{8\pi},
\eqe
where $N'$ is the pair density within the sheet. The synchrotron cooling time is very short at the inferred parameters. Therefore, within the sheet, the reconnection energy release is balanced by the synchrotron cooling. The energy release per unit square of the sheet is determined by the Poynting flux into the sheet, $cE'B'_{\rm pulse}/4\pi$. Introducing the reconnection rate, $\epsilon=E'/B'_{\rm pulse}\sim 0.1$, the energy balance may be written as:
\eqb
\epsilon\frac{B'^2_{\rm pulse}}{4\pi}c=N'\sigma_T\frac{B'^2_{\rm pulse}}{4\pi}c\left(\frac{\varepsilon_T}{m_ec^2}\right)^2a',
\eqe
where $\sigma_T$ is the Thomson cross-section. Eliminating from the last three equations
$\varepsilon_T$ and $n$ in favor of $a$, we obtain:
\eqb
a'=\left(\frac{\epsilon\zeta}{r_e}\right)^{1/2}\left(\frac c{\omega'_B}
\right)^{3/2},
\eqe
where $r_e$ is the classical electron radius and $\omega'_B=eB'_{\rm pulse}/m_ec$ is the cyclotron frequency. 

The emitted frequency in the observer's frame may be estimated by substituting the obtained width of the sheet into Eq. (\ref{freq_reconn}) and making a Lorentz transform. The Lorentz factor of the plasma within the pulse with respect to the wind is given by Eq. (\ref{Gamma}). Just beyond the light cylinder, the wind is only mildly relativistic; therefore, we can use this estimate as the Lorentz factor in the observer's frame. Then, we obtain:
\eqb
\nu=\Gamma'\frac{\omega'}{2\pi}=\frac 1{2\pi\xi} \left(\frac{r_e}{\epsilon\zeta
c\Gamma'}\right)^{1/2}\omega_B^{3/2}
=3\frac{\mu_{33}^{1/4}L_{\rm pulse,47}^{5/8}}{\xi_1\zeta_1^{1/2}\epsilon_{-1}^{1/2}P^2}\,\rm
GHz.
 \label{frequency}\eqe
One sees that the basic parameters of the outgoing radiation are compatible with the observed properties of typical FRBs.

The emitted fms waves are formed within the magnetic pulse, which moves with the relativistic velocity. Therefore, they propagate within the pulse and escape from it far away from the magnetar, when the pulse is decelerated, colliding with the surrounding plasma. Importantly, these waves are polarized perpendicular to the background magnetic field. The magnetic pulse is formed well within
the magnetosphere, and the magnetic flux from this region is transferred by the pulse outward. Therefore, the direction of the magnetic field in the pulse is determined by the rotational phase of the magnetar. When the pulse travels in the magnetar wind, it picks up the azimuthal magnetic field, which is accumulated at the front of the pulse. The radiation passes through this layer on the way out. If the field varies gradually, the polarization is adiabatically adjusted to the local magnetic field, so that the outgoing radiation eventually becomes completely polarized perpendicular to the field in the wind, i.e., along the rotational axis of the magnetar. If the field varies sharply or the accumulated flux is too small, the waves are split into two normal modes corresponding to the local magnetic field so that the radiation is depolarized, whereas the position angle depends on the details of the transition zone. 

In contrast to the synchrotron maser, the reconnection emission model could explain the observed \citep{Gijjar18,Michili18,oslowski19} 100\% degree of linear polarization in a few FRBs. The position angle in this model may be determined by both the field in the magnetar wind and by the magnetosheric field, depending on how the magnetic pulse propagates through the wind. This is compatible with the diverse polarization patterns observed in FRBs \citep{Luo_etal20}. However, the model could not be applied to weak FRBs because, according to Eq. (\ref{frequency}), the emitted frequency becomes too low for small luminosities. This is because the magnetic field in weak pulses is too low already at light cylinder distances. It was suggested \citep{Yuan_etal20} that in the course of the magnetar flare, the reconnection emission could be generated in the upper magnetosphere inside the light cylinder. The ejected plasmoids push out the magnetospheric field lines, so current sheets are formed behind the plasmoid separating oppositely-directed fields. The reconnection in these current sheets could produce the radio emission. A more detailed analysis of this idea is necessary. 

\section{Non-Linear Effects and Escape of Radio Emission}

The high brightness temperature of FRBs implies that the non-linear processes could significantly affect the properties of the outgoing radiation and even prevent escape of radio waves. Here, I briefly outline the relevant processes.

\subsection{Electron in a Strong Electromagnetic Wave}

Let us first consider non-linear effects in the non-magnetized plasma. The strength of the electromagnetic wave is measured by the parameter $a$ introduced in Section 2.5 in Eq. (\ref{strength}). This parameter represents the four-velocity of electron oscillations in the field of the wave (see, e.g., \citep{Melrose_book80}). 
In FRBs, the strength parameter exceeds unity at the distance from the source \citep{Luan_Goldreich14}:
\begin{equation}
 R<2\times 10^{13}\frac{F^{1/2}_{\nu,\rm Jy}D_{\rm Gpc}}{\nu^{1/2}_{\rm GHz}}\,\rm cm.
\end{equation}
The standard linear theory of electromagnetic waves in a plasma assumes that $a\ll 1$. Even under this condition, the non-linear effects could significantly affect the propagation of the wave because even small non-linear corrections accumulate considerably at a long enough path. At $a>1$, the electron oscillates relativistically so that the Lorentz force, $(1/c)\mathbf{v\times B}$, becomes comparable with the electric force. In a linearly-polarized strong wave, the electrons oscillate both in the transverse and in the longitudinal directions, producing a figure-of-eight trajectory in the oscillation-center frame. 

If an electron at rest is illuminated by a strong wave, it is pushed forward with the Lorentz factor $\sim a$. Together with oscillations with amplitude $\sim a$ in the oscillation-center frame, this produces the total energy $\sim m_ec^2a^2$. Therefore, the electron-positron plasma is boosted forward by a strong wave. Conversely, electrons in an electron-ion plasma are tied electrostatically to ions. Therefore, they remain, on average, at rest and only experience oscillations with the amplitude of $\sim a$ \citep{Waltz_Manley78,Sprangle_etal90}.

In many cases, an electron in a high frequency electromagnetic wave may be considered a particle with the effective mass:
\begin{equation}
 m_{\rm eff}=m_e\sqrt{1+\frac 12a^2},
\label{eff_mass}\end{equation}
moving with the velocity of the electron oscillation center, $\mathbf{v}_{\rm d}$, and having the energy and momentum $\varepsilon=m_{\rm eff}c^2\gamma_{\rm d}$ and $\mathbf{P}=m_{\rm eff}\mathbf{v}_{\rm d}\gamma_{\rm d}$, respectively, where $\gamma_{\rm d}=(1-v_{\rm d}^2/c^2)^{-1/2}$ is the Lorentz factor of the oscillation center (see, e.g., \citep{Melrose_book80}). In particular, the conservation laws for the Compton scattering of a strong wave look like
\begin{equation}
 \varepsilon+s\hbar\omega=\varepsilon'+\hbar\omega';\qquad \mathbf{P}+s\hbar\mathbf{k}=\mathbf{P'}+\hbar\mathbf{k'},
\end{equation}
where $s$ is an integer. The scattering of a strong wave may be described
as the absorption of $s$ photons and the reemission of a single photon. Neglecting recoil, we recover the classical picture: a relativistically oscillating electron emits at harmonics of the oscillation frequency. In the regime $a\gg 1$, the power of the scattered radiation is dominated by the higher harmonics of the incident
wave. As with synchrotron emission, the maximal power is achieved at $s\sim a^3$, and the scattering cross-section exceeds the Thomson cross-section $a^2$ times.

The amplitude-dependent relativistic mass of electrons has many implications. The plasma refraction index,
\begin{equation}
 n=\sqrt{1-\frac{\omega_p^2}{\omega^2}},
\label{refraction}\end{equation}
depends on the plasma frequency, 
\begin{equation}
 \omega_p=\sqrt{\frac{4\pi e^2 N}{m_e}}.
\label{omega_p}\end{equation}
Increasing the effective electron mass decreases the plasma frequency. Therefore, the cutoff frequency, below which the wave could not propagate, decreases from $\omega_{\rm cutoff}=\omega_p$ to $\omega_{\rm cutoff}\sim\omega_p/\sqrt{a}$. The dispersion measure decreases appropriately, which could have implications for FRBs \citep{Lu_Phinney20,Yang_Zhang20}. An important point is that these relations are written in the plasma frame, which is the oscillation center frame of electrons.

The synchrotron absorption of strong waves occurs as if the absorbing particles have mass (\ref{eff_mass}) and rotate in the magnetic field with the Lorentz factor $\gamma_{\rm d}$ \citep{Lyubarsky18}.

In some cases, the results of linear theory are strongly modified even at $a<1$. For example, the free-free absorption is determined by the velocity of electrons.
Therefore, if the velocity of oscillations in the wave exceeds the thermal electron velocity, $a>v_T/c$,
the absorption coefficient is suppressed by the factor $(ac/v_T)^3=a^3(m_ec^2/k_BT)^{3/2}$ \citep{Lu_Phinney20}.

All the above effects are based on strong oscillations of electrons in the field of the wave.  If the plasma is magnetized such that the Larmor frequency exceeds the wave frequency, these effects are suppressed for the waves polarized perpendicularly to the background magnetic field \citep{Lyutikov20}; these waves represent the so called X-mode. In the O-mode, where the polarization vector has a component along the background magnetic field, the electron oscillates strongly. Therefore the refraction indexes for the two modes are different, which has important implications for the polarization transfer. When the radiation propagates through a slowly varying magnetic field, the polarization of each mode is adjusted to the local direction of the field if the wavelength  of beating between the two modes is small as compared with the inhomogeneity scale. The polarization of the outgoing radiation is fixed at the so called limiting polarization radius, where this condition is violated (e.g., \citep{Ginzburg70,Cheng_Ruderman79}). For strong waves, these effects were studied in ref.\ \citep{Lu_etal19}. 

\subsection{Induced Compton Scattering}

The induced Compton scattering may be described in terms of radiation transfer theory. 
The scattering process could be considered as absorption followed by emission. Therefore, to consider the induced scattering, we must only multiply, according to the general quantum prescription, the scattering rate by $1+n_{\nu,\mathbf{l}}$, where $n_{\nu,\mathbf{l}}$ is the occupation number of the scattered photons with frequency $\nu$ propagating along the unit vector $\mathbf{l}$. The photon occupation number is simply related to the brightness temperature, $n_{\nu,\mathbf{l}}=k_BT_b/h\nu$. The scattering rate from state $(\nu',\mathbf{l'})$ to state $(\nu,\mathbf{l})$ is proportional to $n_{\nu',\mathbf{l'}}(1+n_{\nu,\mathbf{l}})$. At first glance, this means that the induced scattering rate dominates when the occupation number exceeds unity. However, the net scattering rate is determined by the difference between the scattering $\mathbf{l}\to\mathbf{l'}$ and the reverse process $\mathbf{l'}\to\mathbf{l}$. Then, the product terms $n_{\nu\mathbf{l}}n_{\nu'\mathbf{l'}}$ cancel out unless we take into account that, due to a small recoil, the frequency of the scattered photon decreases in the electron rest frame:
\begin{equation}
 \delta\nu=\frac{h\nu^2}{m_ec^2}(1-\mathbf{l\cdot l'}).
\end{equation}
Let us consider photons with the frequency $\nu$ propagating into the direction $\mathbf{l}$ and assume for a while that they are scattered only into or from the direction $\mathbf{l}'$. 
The photon scattered from the direction $\mathbf{l}'$ to the direction $\mathbf{l}$ has the frequency $\nu+\delta\nu$ whereas when a photon is scattered from the direction $\mathbf{l}$ into the direction $\mathbf{l}'$, the frequency of the scattered photon is $\nu-\delta\nu$. Therefore the net rate of scattering into and out of the state $(\nu,\mathbf{l})$ is proportional to
\begin{equation}
 n_{\nu+\delta\nu,\mathbf{l'}}\left(1+n_{\nu\mathbf{l}}\right)\nu^2-n_{\nu\mathbf{l}}\left(1+n_{\nu-\delta\nu,\mathbf{l'}}\right)(\nu-\delta\nu)^2=(n_{\nu\mathbf{l'}}-n_{\nu\mathbf{l}})\nu^2+2n_{\nu\mathbf{l}}\frac{\partial \nu^2n_{\nu\mathbf{l'}}}{\partial\nu}\delta\nu,
\end{equation}
where the factor $\nu^2$ takes into account the phase volume. The induced scattering rate exceeds the spontaneous scattering rate only when the second term in the rhs of this equation exceeds the first term, i.e., if $n_{\nu\mathbf{l}}\delta\nu>\nu$. This requires a very high brightness temperature, $k_BT_b(1-\mathbf{l\cdot l'})\gg m_ec^2=k_B\times 6\cdot 10^9$ K.

The total scattering rate is  found by multiplying the above expression by the scattering probability and integrating over the directions $\mathbf{l}'$. When the plasma is nonrelativistic, $v_T\ll c$, and the radiation spectrum is wide, $\Delta\nu/\nu\gg v_T/c$, the resulting radiation transfer equation for the induced scattering is written as (e.g., \citep{Wilson82}):
\eqb
\frac{d I_{\nu,\mathbf{l}}}{dt}
=\frac{r_e^2Nc}{m_e}I_{\nu,\mathbf{l}}
 \int(\mathbf{e\cdot e}')^2
(1-\mathbf{l\cdot l}')\frac{\partial
}{\partial\nu}\left(\frac{I_{\nu,\mathbf{l'}}}{\nu}\right)d\Omega',
 \label{kinComp}\eqe
where $I_{\nu,\mathbf{l}}=h\nu^3n_{\nu,\mathbf{l}}/c^2$ is the radiation intensity, $N$ is the electron number density, and $\mathbf{e}$ is the
polarization unit vector. The time derivative is along the ray.
The induced scattering does not affect the escape time of photons from the source, but redistributes them toward lower frequencies, thus heating the plasma \citep{Syunyaev71}. The total number of photons, $\int\omega^2 n_{\nu,\mathbf{k}}d\omega d\Omega$, is conserved. If the initial spectrum has a maximum, such that $I_{\nu,\mathbf{l}}/\nu$ increases with $\nu$ at $\nu<\nu_0$ and decreases at $\nu>\nu_0$, the induced scattering increases the intensity at $\nu<\nu_0$ and decreases it at $\nu>\nu_0$, the maximum intensity shifting toward larger frequencies.

Note that the equation does not contain the Planck constant, even though the above simple picture is based on the quantum approach. The reason is that the effect is purely classical: the non-linear interaction of two waves yields a beating wave, 
which exerts a constant force on electrons moving with the velocity equal to the beating phase velocity, $\omega-\omega'=\mathbf{v\cdot(k-k')}$. This transfers the energy from the wave to the plasma, which means that the wave with the higher frequency decays. The classical derivation of the induced Compton scattering of electro-magnetic waves is given in \citep{Galeev_sunyaev73,Drake_etal74}.

The probability of the induced scattering into a state is proportional to the number of photons already available in the state. The radiation transfer equation (\ref{kinComp}) has the form:
\begin{equation}
 \frac{d I_{\nu,\mathbf{l}}}{dt}
=GI_{\nu,\mathbf{l}},\qquad G==\frac{r_e^2Nc}{m_e}
 \int(\mathbf{e\cdot e}')^2
(1-\mathbf{l\cdot l}')\frac{\partial
}{\partial\nu}\left(\frac{I_{\nu,\mathbf{l'}}}{\nu}\right)d\Omega',
\label{ind_rate1}\end{equation}
so that the quantity $G$, which depends on the radiation intensity, is the induced scattering rate. 
Within the source, where the radiation is nearly isotropic, the recoil factor $(1-\mathbf{l\cdot l}')$ is of the order of unity. Then, we can estimate the scattering rate as $G\sim (k_BT_b/m_ec^2)\sigma_Tnc$, where $\sigma_T$ is the Thomson cross-section. If the plasma in the source is relativistically hot, the induced scattering rate decreases by $\gamma_T$, which is the characteristic thermal Lorentz factor of electrons in the source \citep{Lyubarsky08}. The escape time from a source of size $L$ is $L/c$. Then, the optical depth to the induced scattering is conveniently defined as the ratio of the escape time to the frequency redistribution time:
\begin{equation}
 \tau_{\rm ind}=GL/c\sim
 \frac{k_BT_b}{m_ec^2}\sigma_TNL.
\end{equation}
The source is transparent to the induced scattering if $\tau_{\rm ind}<1$. In the opposite case, the radiation is redistributed toward smaller frequencies before escaping.

If the radiation is highly beamed, as is the case at large distances from the source, the recoil factor $(1-\mathbf{l\cdot l'})$ makes the scattering within the beam inefficient. Then, the scattering outside the beam dominates \citep{Coppi_etal93,Lyubarsky08} because, according to Equation (\ref{ind_rate1}), even weak isotropic background radiation (created, e.g., by spontaneous scattering) grows exponentially. Let us choose in Eq. (\ref{ind_rate1}) $\mathbf{l'}$ in the direction of the beam and $\mathbf{l}$ in the direction well outside the beam, where only a weak background emission is initially present.
Then, the intensity in the direction $\mathbf{l}$ grows exponentially with the rate:
\begin{equation}
 G\sim \sigma_TNc\frac{F_{\nu}}{m_e\nu^2},
\label{ind_rate}\end{equation}
where $F_{\nu}= I_{\nu,\mathbf{l}}\Delta\Omega$ is the spectral flux at the scattering point and $\Delta\Omega$ is the solid angle subtended by radiation at this point. Eventually, the scattered radiation takes whole the energy of the primary beam. Inasmuch as the intensity in the direction $\mathbf{l}$ is initially very low, many e-folding times are necessary to remove significant energy from the beam. Therefore, in this case, the transparency condition may be written as:
\begin{equation}
 \tau_{\rm ind}= \frac 1c\int GdR\sim GR/c<10.
\label{transp}\end{equation}
Here, the flux, and therefore the rate $G$, decreases as $R^{-2}$; therefore, only the region $\Delta R\sim R$ contributes into the integral.

The above assumes that the emission is steady. In the case of a short pulse with duration $\tau<R/c$, the induced scattering occurs
only while the scattered ray remains within the zone illuminated by the primary radiation. Therefore, the optical depth is determined not by the size of the medium but by the duration of the pulse, so that we must substitute $R$ by $c\tau$ in Eq. (\ref{transp}) \citep{Lyubarsky08}. 

The above considerations also assume that the electrons oscillate non-relativistically in the field of the wave. This means that the wave strength parameter (Eq. (\ref{strength}) is smaller than unity. With this parameter, the induced scattering rate (\ref{ind_rate}) may be presented as:
\begin{equation}
 G\sim\frac{\omega_p^2a^2}{\omega},
\end{equation}
where $\omega_p$ is the plasma frequency (\ref{omega_p}).
When $a>1$, the scattering rate decreases because the effective mass of a relativistically oscillating electron increases. A larger effective mass reduces recoil, which is crucial for the induced scattering. In this case, the scattering rate is estimated as \citep{Lyubarsky19a}:
\begin{equation}
 G\sim\frac{\omega_p^2}{\omega a}.
\end{equation}
Even though the spontaneous scattering is dominated by high harmonics at $a\gg 1$, the induced scattering occurs predominantly in the first harmonic at any $a$.

In FRBs, the induced scattering could affect the wave propagation close enough to the source. In synchrotron maser models (see Section 3.3), the transparency condition (\ref{transp}) is marginally satisfied for $\sim 1$ GHz and violated for smaller frequencies \citep{Metzger19,Beloborodov20}, which could explain the non-detection of low frequency FRBs \citep{Karastergiou15}. The effect of the induced scattering was also used in order to show that FRBs could not be produced in stellar coronae \citep{Lyubarsky_Ostrovska16}.

\subsection{Induced Raman Scattering}

Raman scattering is the scattering of electromagnetic waves on plasma oscillations. The process may be interpreted as a decay of an electromagnetic wave into another electromagnetic wave and a plasma wave or/and a merging of an electromagnetic and a plasma waves into another electromagnetic wave. The energy and momentum conservations imply relations between the frequencies and the wave vectors of three waves:
\begin{equation}
 \omega=\omega'+\omega_p;\qquad \mathbf{k=k'+q},
\label{resonance}\end{equation}
where $\omega$, $\mathbf{k}$, and $\omega'$ and $\mathbf{k'}$ are the frequencies and wave vectors of the electromagnetic waves, respectively; $\omega_p$ is the plasma frequency (\ref{omega_p}); and $\mathbf{q}$ is the wave vector of the plasma wave. If a powerful beam propagates through a medium with no preexisting plasma turbulence, the induced Raman scattering results in the exponential growth of both the intensity of the scattered wave and the level of plasma turbulence (parametric instability \citep{Drake_etal74}). With the classical language, the non-linear interaction between two waves produces a beating wave, which resonantly excites the third wave.

Importantly, this process is impossible in electron-positron plasma. The reason is that the beating wave exerts the same force both on electrons and on positrons, so that the plasma waves in which electrons and positrons oscillate in antiphase are not excited. 

In electron-ion plasma, the rate of the induced Raman scattering could be represented as \citep{Thompson94,Lyubarsky08}:
\begin{equation}
 G=\sigma_TNc\frac{F_{\nu}}{m_e\nu_p\nu}(1-\cos\theta),
\label{rate1}\end{equation}
where $\theta$ is the scattering angle. If $\theta$ is not small, the Raman scattering rate exceeds that of Compton scattering (\ref{ind_rate}). However, due to Landau damping, the phase velocity of the plasma wave could not be smaller than a few electron thermal velocities, which may be written as $q<\omega_p/4v_T$. Therefore, the scattering angle is limited by the resonance conditions (\ref{resonance}), so that the Raman scattering rate exceeds the Compton scattering only if:
\begin{equation}
 \frac{\omega}{\omega_p}<\frac{m_ec^2}{8k_BT}.
\end{equation}
The condition of efficient scattering is still given by Eq. (\ref{transp}) because, just as for the case of the induced Compton scattering, many e-folding times are necessary for the scattered waves to grow significantly.

Importantly, the scattering by a small angle does not prevent escape of radiation, it only makes the beam wider, which could smear a short pulse. However, the situation with short pulses is more complicated so that we cannot only substitute $R$ by $c\tau$ as with induced Compton scattering. As for the case of induced Compton scattering, the process occurs only in the illuminated region. However, if the scattering angle is small, the scattered wave remains within the illuminated region along a large distance, $L=c\tau/(1-\cos\theta)$. Convesely, the group velocity of the plasma waves is small; therefore, they are amplified only for the time $\tau$. The Raman scattering of short pulses was addressed considering these effects in \citep{Lyubarsky08}.

Notably, the scattering rate (\ref{rate1}) was obtained by neglecting the decay of the plasma waves, i.e., assuming that the decay rate is less than the scattering rate. If this condition is violated, the Raman scattering is suppressed.
The collisional decay of plasma waves could be easily taken into account by comparing the decay rate with the scattering rate \citep{Thompson94,Lyubarsky08}. Since the latter is proportional to the beam intensity, the collisional decay of plasma waves may be neglected for powerful enough radiation beams. However, the Raman scattering of powerful beams could be affected by the non-linear interactions of plasma waves, such as the induced scattering of plasma waves (non-linear Landau damping) and the modulation instability (see, e.g., \citep{Shapiro_Shevchenko84,BreizmanREVIEW}). This could prevent the growth of plasma turbulence beyond some level, thus suppressing the rate of Raman scattering. To determine the role of the Raman scattering in any particular case, one must consider all the above effects.

\subsection{Modulation and Filamentation Instabilities}

Due to the non-linear interactions of radiation and matter, the refraction index depends on the radiation power, which could lead, under some conditions, to self-modulation or/and self-focusing of the radiation beam (see, e.g., \citep{Karpman75}). Let the refraction index increase with the radiation intensity. Then, any transverse intensity gradient is amplified because the rays are focused toward the intensity maximum. In wide beams, self-focusing leads to filamentation, i.e., to modulation in the direction perpendicular to the propagation direction.
Self-modulation in the longitudinal direction occurs if the wave group velocity decreases with intensity. Then, the waves accumulate near the intensity maximum. 

In plasma, the refraction index is given by Eq. (\ref{refraction}).
Three effects provide the dependence of the refraction index on the radiation intensity and, therefore, could potentially lead to modulation and/or filamentation instabilities:

The first effect is based on the ponderomotive force, which expels plasma from the regions of enhanced radiation intensity. Therefore, if this fluctuation forms, the plasma frequency in the region decreases so that the refraction index increases. Due to refraction, the radiation is focused at the region, thus repelling more plasma and amplifying the focusing effect. This effect leads to filamentation of the beam; however, it could not produce self-modulation \citep{Shearer_Eddleman73,Kaw_etal73,Drake_etal74}. The effect works efficiently in electron-positron plasma, whereas in electron-ion plasma, it is suppressed by the large inertia of ions, and therefore it is only involved at very large intensities.

The second effect reduces the plasma frequency in the regions of enhanced radiation because, as discussed in Section 4.1, the effective mass of the oscillating electron increases with oscillation amplitude. This results in increasing the refraction index. 

The third effect arises because, in the field of the wave, the longitudinal $\mathbf{v\times B}$ force produces an electron density perturbation, $\Delta n$, at the double wave frequency. The beating between this perturbation and the electron velocity oscillations in the pumping wave yields current $\mathbf{j}=\delta N\mathbf{v}$ at the wave frequency, which affects the propagation of the pumping wave. The last two effects are not suppressed by the ion inertia because they do not affect the average plasma density. They are of the same order, and therefore should be considered together. It is shown \citep{Max_etal74} that they lead to both filamentation and modulation of the radiation beam. The scale of filamentation is significantly larger than the longitudinal modulation scale, so the beam breaks into pancakes transverse to the radial direction.

Thus, in the electron-positron plasma, where the effect of the ponderomotive force dominates, only the filamentation instability could develop. In particular, simulations of the maser emission from relativistic shocks reveal filamentation of the upstream flow \citep{Iwamoto_etal17}. In electron-ion plasma, the effect of the ponderomotive force is suppressed by ion inertia; therefore, in a wide range of parameters, both filamentation and modulation are possible.  The comprehensive analysis of different regimes in electron-ion plasma was given in \citep{Sobacchi_etal20}.

Neither filamentation nor modulation prevent escape of radiation from the source. However, they could affect the properties of the outgoing radiation. It was suggested \citep{Yang_Zhang20,Lyutikov20} that self-focusing could affect the collimation angle and the true event rate of FRBs. However, when the beam is macroscopically wide, it is not focused as a whole. Instead, the filamentation instability develops faster on small scales; therefore, the beam only breaks into narrow subbeams, each of which may be also modulated in the longitudinal direction. This could produce other observable effects \citep{Sobacchi_etal20}. Namely, when the burst exits the plasma slab, the subbeams are strongly diffracted, which could lead to (1) smearing of the burst in time if the diffraction angle is large enough and (2) frequency modulation of the observed intensity due to interference between the subbeams. 
The longitudinal modulation could, in turn, imprint a microsecond structure on the light curve of the burst. 

\subsection{Non-Linear Interaction of Waves in a highly magnetized plasma}

Let us consider the most relevant for the magnetar magnetospheres case when the wave frequency is small compared with both the plasma and the Larmor frequencies and when the magnetic energy considerably exceeds the plasma energy. Then, we can work in the scope of the force-free MHD. Only two types of waves exist in this system: the fast magnetosonic (fms) and Alfven waves. The fms waves were briefly discussed in Section 4. Within the limit $\sigma\to\infty$, their dispersion law is reduced to $\omega=ck$. The dispersion law of the Alfven waves in the same limit is $\omega=ck\vert\cos\theta\vert $, where $\theta$ is the angle between the background magnetic field and the propagation direction of the wave. The fms waves are a low frequency $\omega\ll\omega_p$ limit of the so called X-mode of electromagnetic wave (e.g., \citep{Arons_Barnard86}). When propagating towards decreasing plasma density, the fms wave is smoothly converted into the X-mode and could, in principle, eventually escape as outgoing electromagnetic waves. The Alfven waves propagate only along the magnetic field lines and therefore if they are excited on closed field lines, they remain trapped. 

The non-linear interactions of force-free MHD waves were studied in refs.\ \citep{Thompson_Blaes98,Lyubarsky20}. The strongest are the resonant three-wave interactions, i.e., the decay of a wave into two waves and the merging of two waves into one. In this case, the resonant conditions (the conservation laws in the quantum language) looks like
\begin{equation}
 \omega=\omega_1+\omega_2;\qquad \mathbf{k=k_1+k_2}.
\end{equation}
Substituting the dispersion laws into the conservation laws, one sees that only interactions
involving both types of waves are possible:
\begin{equation}
 {\rm fms}\longleftrightarrow {\rm fms}+{\rm Alfven}\quad {\rm and}\quad
{\rm fms}\longleftrightarrow {\rm Alfven}+{\rm Alfven}.
\end{equation}
 The rate of the interaction may be roughly estimated as:
\eqb
G\sim\left(\frac{\delta B}{B}\right)^2\omega,
 \label{q}\eqe
where $\delta B$ is the amplitude of the waves and $B$ is the background field.

Assume that there is a powerful source of fms waves well within the magnetar magnetosphere. As they propagate outwards, their amplitude decreases with the distance as $1/R$, whereas the background field in the magnetosphere decreases as $1/R^3$, so that:
 \eqb
\frac{\delta B}B=\sqrt{\frac{L_{\rm FRB}}c}\frac{R^2}{\mu}.
 \label{relativ-ampl}\eqe
Therefore, the characteristic time of the non-linear interaction, $1/G$, rapidly decreases with distance. For waves in the GHz band, it becomes smaller than the propagation time, $R/c$, already at distance \citep{Lyubarsky20}:
\eqb
R= 10^7\left(\frac{\mu_{33}^2}{L_{\rm FRB, 43}\nu_9}\right)^{1/5}\,\rm cm,
 \eqe
which is well within the magnetosphere. 
Beyond this distance, the non-linear interaction yields a cascade redistribution of the waves toward larger and smaller frequencies. High-frequency waves eventually decay. The amplitude of low-frequency waves increases with decreasing frequency because of energy conservation. Therefore, eventually, the non-linear steepening is involved and they decay too. 
As no other waves in the GHz band can propagate across magnetic field lines, even if a source of GHz waves with the required power existed inside the magnetosphere, no radiation would escape. 

In the outer magnetosphere, the situation is different. The FRB is triggered by the magnetar flare, which produces a magnetic perturbation (the magnetic pulse) described in Section 3.2. In the outer magnetosphere, the field of the pulse is larger than the local magnetospheric field, and the pulse propagates with relativistic velocity. When the pulse excites high-frequency waves, as described in Section 3.4, they propagate on the top of the pulse to very large distances. They are not strongly affected by non-linear interactions because the ratio of the wave amplitude to the field in the pulse is small and because of relativistic  time dilation within the pulse \citep{Lyubarsky20}. 

\section{Conclusions}

In this review I outlined 
the radiation mechanisms proposed for FRBs. The aim was to demystify, at least somewhat, the field of the coherent emission mechanisms. I tried to describe them at the very basic physics level, avoiding technical details. 
However, one has to stress again, that these mechanisms may be understood only as collective plasma processes; an over-simplistic approach could not provide reasonable results. 

For a while, we have two workable mechanisms: the synchrotron maser at the front of a relativistic magnetized shock and the radiation from variable currents in a reconnecting current sheet. Both mechanisms assume relativistic magnetized outflows, which could be found in different astrophysical environments. Observational evidence as well as theoretical considerations point to magnetar flares as the most promising progenitor. Therefore, both types of FRB models were developed within the scope of the magnetar paradigm. Both models have their pros and cons. The synchrotron maser produces radio emission in a wide range of luminosities; however, this emission could hardly be 100\% polarized as is sometimes observed. Moreover, shock waves with the required parameters may be obtained only with an assumption that a moderately relativistic and mildly magnetized medium is present relatively close to the magnetar. The origin of such a medium is unclear. The emission from a reconnecting current sheet may have a 100\% degree of polarization; however, this model could not explain weak FRBs because the emission frequency becomes too low for small luminosities. This emission has been studied only for a quiet, steady reconnection. It is unclear how the process develops in a highly unsteady and turbulent reconnection when the sheet is destroyed 
by a strong electromagnetic perturbation caused by the magnetar flare. Further development of both models is necessary. It is possible that both mechanisms work in different cases. And of course we could not exclude that new models will be presented that will successfully agree with observations.

\acknowledgments{I gratefully acknowledge grant I-1362-303.7/2016 from the German-Israeli Foundation for
Scientific Research and Development and grant 2067/19 from the Israeli Science Foundation.}

\reftitle{References}

\externalbibliography{yes}
\bibliography{FRB}

\end{document}